\documentclass[prl,twocolumn,showpacs,amssymb]{revtex4}

\usepackage{graphicx}
\usepackage{bm}
\def\be{\begin{equation}}
\def\ee{\end{equation}}
\def\bea{\begin{eqnarray}}
\def\eea{\end{eqnarray}}

\begin{document}
\title{Nonequilibrium kinetics of a disordered Luttinger liquid}
\author{D. A. Bagrets$^{1}$}
\author{I. V. Gornyi$^{1,2}$}
\author{D. G. Polyakov$^{1}$}
\affiliation{
$^{1}$Institut f\"ur Nanotechnologie, Forschungszentrum Karlsruhe,
76021 Karlsruhe, Germany
\\
$^{2}$A.F.Ioffe Physico-Technical Institute,
194021 St.~Petersburg, Russia
}
\date{\today}

\begin{abstract}
We develop a kinetic theory for strongly correlated disordered one-dimensional
electron systems out of equilibrium, within the Luttinger liquid model. In the
absence of inhomogeneities, the model exhibits no relaxation to equilibrium. We
derive kinetic equations for electron and plasmon distribution functions in the
presence of impurities and calculate the equilibration rate $\gamma_E$.
Remarkably, for not too low temperature and bias voltage, $\gamma_E$ is given by
the elastic backscattering rate, independent of the strength of electron-electron
interaction, temperature, and bias.
\end{abstract}

\maketitle

{\it Introduction.}---Interacting electrons in one dimension (1D)
\cite{giamarchi04} have become a focus of interest in nanophysics from both the
fundamental and applied perspectives. Recent technological advances have made it
possible to systematically study the transport properties of a variety of
ultranarrow wires; in particular, carbon nanotubes \cite{man05} and semiconductor
nanowires \cite{auslaender05}. From the fundamental point of view, much of the
fascination with physics of 1D systems is driven by the fact that
electron-electron (e-e) interactions in 1D geometry are qualitatively significant,
transforming the electron gas into a Luttinger liquid (LL) \cite{giamarchi04}.
This strongly correlated state of matter is commonly described in terms of bosonic
elementary excitations.

A conceptually nontrivial aspect of the non-Fermi liquid nature of a LL concerns
its behavior at nonequilibrium, e.g., when a finite bias voltage is applied to the
wire (for a recent experiment see Ref.~\cite{chen09}). A homogeneous LL is
completely integrable and as such does not exhibit any relaxation to equilibrium:
an excited state will never decay to the state characterized by temperature $T$.
Quite remarkably, a finite quantum lifetime of fermionic excitations due to e-e
interactions in a homogeneous LL does not translate into any {\it inelastic} e-e
scattering \cite{gornyi05}: the allowed energy transfer is exactly zero. This is
in stark contrast to electron liquids in higher dimensions, where the
characteristic energy transfer is $T$. Relaxation to equilibrium due to e-e
collisions in a LL is thus only possible if momentum conservation is broken by
inhomogeneities. Of central importance is therefore the question---essentially
unanswered---of how the equilibration in a LL occurs in the presence of a random
backscattering potential. This is the subject of this work.

So far, advances in dealing with a LL off equilibrium have been focused on a
``mechanical" approach, i.e., on solving the equations of motion ``as exactly as
possible". Efforts, based on the bosonization approach, have been centered around
the nonlinear conductance of a LL containing a single compact scatterer
\cite{furusaki93}. %fendley95,egger98}.
However, the line of research relying on the
exact integrability cannot possibly be much extended beyond the single-scatterer
case (for two tunneling barriers in a LL, a nonequilibrium distribution of
plasmons was studied by means of the master equation, see
Ref.~\cite{kim03}). Also, importantly, the conventional bosonization
\cite{giamarchi04} is designed for equilibrium boundary conditions. An alternative
is to preserve dynamics of both bosonic and fermionic degrees of freedom
\cite{bagrets08,gutman08}, as it is done in the {\it functional} bosonization
approach (see Ref.~\cite{yurkevich04} for a review).

Our purpose here is to develop a {\it kinetic} approach to nonequilibrium
phenomena in a disordered LL, by formulating kinetic equations for {\it
distribution functions}. Within this approach, one {\it has} to introduce the
distribution functions of not only bosonic but also fermionic excitations,
similarly to higher dimensions \cite{catelani05,remark_ghost}. Our main result is
a set of kinetic equations which describe (i) inelastic e-e scattering, mediated
by virtual plasmons, and (ii) creation/annihilation of real plasmons; both
processes being only triggered by scattering off disorder. We calculate a key
quantity in nonequilibrium problems: the energy relaxation rate $\gamma_E$. In a
remarkable departure from Fermi liquids, $\gamma_E$ in a LL at not too low $T$
turns out to be given by the {\it elastic} scattering rate $\gamma$.

{\it Effective action.}---We study interacting electrons in a single-channel
disordered quantum wire within the LL model \cite{giamarchi04}: the electron
dispersion relation is linear (with the velocity $v_F$) and interactions yield
only forward e-e scattering (characterized by the dimensionless constant
$\alpha=V_{\rm f}/2\pi v_F$, where $V_{\rm f}$ is the zero-momentum Fourier
component of the interaction potential). We consider both spinless ($\eta = 1$)
and spinful ($\eta = 2$) models. Our approach is based on the ``quasiclassical"
real-time electron Green's function at coinciding spatial points ${\hat
g}(x,t_1,t_2)$ \cite{rammer86}, widely used in the nonequilibrium theory of
mesoscopic transport. The ``hat" means that $\hat g$ is a matrix in the Keldysh,
chirality ($\mu=\pm$ for right/left-moving electrons) and (possibly) spin
($s=\uparrow, \downarrow $) spaces. In what follows we use the Pauli matrices
$\tau$, $\sigma$, and $s$ that act in the chirality, Keldysh, and spin spaces,
respectively.

The quasiclassical Green's function satisfies the condition ${\hat g}\circ {\hat
g}=\delta (t_1-t_2)$, where the ``dot" denotes the convolution in the full
(chirality$\times$spin$\times$Keldysh$\times$time) space. This constraint enables
us to describe the wire by the action that reproduces the averaged over disorder
equation of motion for $\hat g$ as its saddle point. The way to derive such an
action is similar to that in the ballistic sigma-model in higher
dimensions~\cite{khmelnitskii95}. To account for the Coulomb interaction, one
introduces the Hubbard-Stratonovich field $\hat \phi(x,t) = \phi_1 +  \sigma_x
\phi_2$ on the Keldysh contour~\cite{kamenev99}, where the ``classical" and
``quantum" fields $\phi_{1,2}$ are diagonal matricies in the chirality and spin
spaces. Then the action takes the form \cite{bagrets08}
\bea
S\{\hat{g},\hat\phi\} &=& {1\over 2}\,{\rm Tr}\int\!dx \left[\,\frac{1}{v_F^*}
\left(-i\partial_t + \hat\phi\right)\tau_z \hat{g}
-  \hat{g}_0{\cal T}^{-1}i\partial_x{\cal T} \right.\nonumber\\
& - &\left.\frac{i\gamma_0}{16v_F^*}\tau_+\hat{g}\tau_-\hat{g} + \hat\phi\,
\left(\frac{\sigma_x}{2\pi v_F^*} + \hat V_0^{-1}\right)\hat\phi\,\right].
\label{ActionK}
\eea
Here $\gamma_0$ is the bare elastic rate of backscattering off a random static
potential, $\hat{g}_0= {\rm diag}\,({g}^+_0,-{g}^-_0)$ corresponds to the saddle
point of the action of the noninteracting problem, and $\hat{g}={\cal
T}\hat{g}_0{\cal T}^{-1}={\rm diag}\,({g}^+,- {g}^-)$. The unitary transformation
$\cal T$ (diagonal in the chirality$\times$spin space) parametrizes fluctuations
around $\hat{g}_0$ due to fluctuations of $\hat\phi(x,t)$. The interaction of
fermions of the same chirality and spin is included in the shift of the Fermi
velocity $v_F^* =  v_F+V_{\rm f}/2\pi$ \cite{gornyi05}. The last term in
Eq.~(\ref{ActionK}) with $\hat V_0^{-1} = V_{\rm f}^{-1}\sigma_x\tau_x\left[1+2
(\eta-1)s_x \right]$ accounts for true interactions. A similar (replicated
imaginary-time) action for a disordered LL at equilibrium was proposed in
Ref.~\cite{micklitz08}.

{\it Quantum kinetic equations.}---Starting from the effective Keldysh action
(\ref{ActionK}), we use the standard procedure \cite{rammer86,zala01} to derive
the kinetic equations. We proceed at one-loop order with respect to the effective
interaction, which is equivalent to the ``dirty random-phase approximation" (dRPA)
\cite{gornyi05}. The one-loop derivation is controlled by the parameters
$\gamma/\max\{T,eU\} \ll 1$ and $\alpha \ll 1$, which is assumed in the rest of
the paper ($U$ is the bias voltage).
We also disregard the localization effects \cite{gornyi05,yashenkin08}. 
For a wire of length $L\agt v_F/\gamma$, this limits the
applicability of what follows to $\max \{T,eU\}\gg T_1=\gamma/\alpha^{3-\eta}$.

Within the dRPA, we expand the action~(\ref{ActionK}) in fast (on a scale of the
relaxation time for the kinetic equation) quadratic fluctuations of $\hat\phi$
around the ``slow" semiclassical saddle point ${g}^\mu_{\rm slow}=
(\sigma_z+\sigma_+)\delta(t_1-t_2)-2\sigma_+f^{\mu}(x,t_1,t_2)$. The electron
distribution function $f^\mu_\epsilon(x,t)$ at given energy $\epsilon$ is defined
via the Wigner-transform of $f^\mu(x,t_1,t_2)$ and is yet to be found from the
kinetic equations. In this way we obtain the Gaussian action with the propagator
$\hat V=(\hat V_0^{-1}-\hat\Pi)^{-1}$. Here $\hat\Pi=\sigma_x (\partial_t{\hat
D}\, \sigma_x-1)\!/\,2\pi v_F^*$ is the polarization operator, with $\hat D$ being
the electron-hole (e-h) propagator damped by disorder. The retarded part of $\hat
D$ is given by $D_R^{-1} = -i(\omega-\tau_z v_F^* q) + \gamma (1-\tau_x)/2$, while
the kinetic part $D_K$ is expressed (see below) in terms of $f^\mu_\epsilon$ via
\be
N^{\mu\nu}_\omega = \frac{1}{2\omega}\int\!d\epsilon \left[ f^{\mu}_{\epsilon}
(1-f^{\nu}_{\epsilon-\omega}) + (\mu \leftrightarrow  \nu) \right].
\label{Neh}
\ee

To derive the kinetic equation for electrons, we average out the fast fluctuations
in the equation of motion for $\hat g(x,t_1, t_2)$ with the dRPA action, which
yields \cite{remark}
\be
(\partial_t+\mu v_F^* \partial_x ) f^{\mu}_\epsilon =
 -\frac{\gamma + \gamma_{\rm inel}}{2}
( f^{\mu}_\epsilon-f^{-\mu}_\epsilon) +
   {\rm St}^{\mu}_{e-b}(\epsilon)~.
\label{Fermions_KE}
\ee
The collision integral ${\rm St}^{\mu}_{e-b}$ describes inelastic electron
scattering due to interaction with the bosonic bath,
\be
{\rm St}^{\mu}_{e-b}\!=\!\sum_\nu \int \!\!d\omega
I^{\mu\nu}(\omega) \left[
f^{\nu}_{\epsilon+\omega}(1-f^{\mu}_\epsilon)
- f^{\mu}_{\epsilon}(1-f^{\nu}_{\epsilon-\omega})
\right], \label{St1}
\ee
while $\gamma_{\rm inel}$ accounts for the additional \cite{remark} renormalization of the static disorder
due to the inelastic scattering,
\be
\frac{\gamma_{\rm inel}}{2} = - \int  d\omega\, I^{+-}(\omega) \left(1 -
f_{\epsilon-\omega} + f_{\epsilon+\omega} \right). \label{St2}
\ee
Here $f_\epsilon=(f^{+}_\epsilon+f^{-}_\epsilon)/2$, and $I^{\mu\nu}(\omega)$ is
the rate of emission of energy per unit interval of $\omega$, accompanied by
scattering $\mu\to\nu$, which is given by
\be
I^{\mu\nu}(\omega)=\frac{i}{\pi}\int\frac{dq}{2\pi} \,V_{>,\parallel }^{\nu\mu}(\omega,q)
\,\,{\rm Re}\,D^{\mu\nu}_R(\omega,q)~. \label{KernK}
\ee

In the integrand (\ref{KernK}), there are four poles, $q\simeq\pm \omega (1 \pm
i\gamma/2\omega)/v_F^*$, inherited from ${\rm Re}\,D_R$, which are only slightly
damped by disorder in the limit $\omega \gg \gamma$. They correspond to e-h pair
excitations described by the renormalized Fermi velocity $v_F^*$. Four more poles
$q\simeq\pm \omega (1 \pm i\gamma/2\omega)/u$, associated with the ``greater" part
of the effective interaction with parallel spins $V_{>,\parallel}$, correspond to
the collective plasmon mode of the clean LL, moving with velocity $u=v_F(1 + \eta
\alpha)^{1/2}$. In the spinful model, $\hat V$ contains an extra
mode---spinon---propagating with velocity $v_F$ \cite{yashenkin08}.
Importantly, the e-h and collective excitations at $\omega \gg T_1$ are well
resolved from each other and should be treated separately, while in the opposite
limit, $\omega \ll T_1$, the disorder-induced quantum uncertainty makes them
indistinguishable. We now proceed by relating $V^{\nu\mu}_{>,\parallel}$ to the
distribution functions.

\begin{figure}[t]
\includegraphics[width=2.9in]{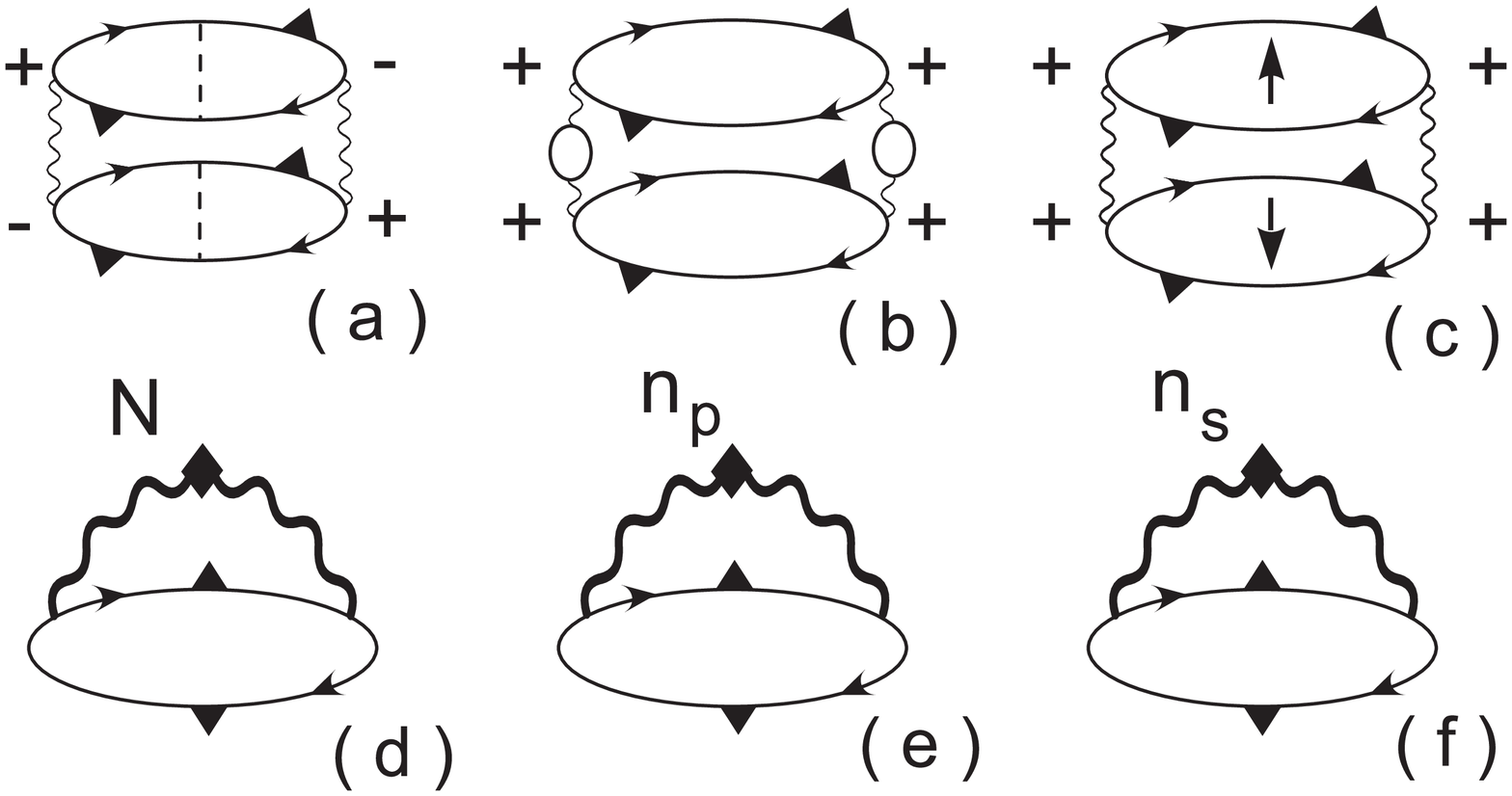}
\caption{Scattering processes corresponding to the collision kernels
${\cal K}(\omega)$ and ${\cal L}(\omega)$. Thin wavy lines: bare interaction $V_{\rm f}$.
Solid wavy lines: dRPA interaction propagator $V_>$. The triangles ($\blacktriangle$) and rhombi
($\blacklozenge$)  mean the Keldysh part of the electron and boson Green's functions, respectively.
}
\vspace{-2mm}
\label{Scattering}
\end{figure}

\begin{figure}[t]
\includegraphics[width=2.6in]{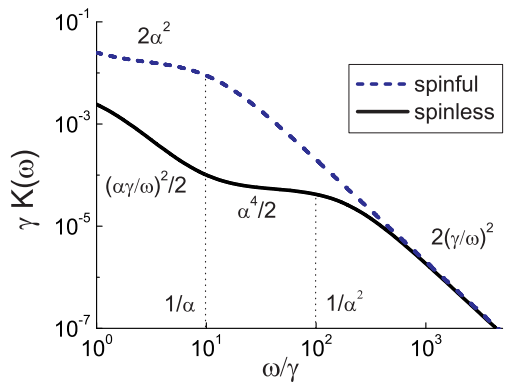}
\caption{
Frequency dependence of the collision kernels
for the spinless and spinful
models [Eq.~(\ref{ColK})] for $\alpha=0.1$.
%[Eq.~(\ref{K}) and~(\ref{Ks}) ].
}
\label{Kw_plot}
\end{figure}

{\it Small energy transfer, $\omega \ll T_1$}.---In this limit, the Keldysh,
retarded, and advanced parts of interaction satisfy $V_K\simeq V_R\Pi_K V_A$. Then
the general result for the emission rate can be represented in the form
$I^{\mu\nu}(\omega) = \sum_{\alpha\beta} \omega\, {\cal
K}^{\mu\nu}_{\alpha\beta}(\omega) N^{\alpha\beta}_\omega$, where the collision kernel
${\cal K}^{\mu\nu}_{\alpha\beta}(\omega)$ describes inelastic spin-conserving
electron scattering $\mu \to\nu $ with energy transfer $\omega$ to
the electron and hole having the chirality $\alpha$ and $\beta$. For the
total collision kernel ${\cal K} = \frac{1}{2}\sum_{\mu\nu} {\cal
K}^{\mu\nu}_{\mu\nu}$ we obtain
\be
{\cal K}(\omega)=\int
\frac{dq}{2\pi^2\omega} \sum_{\mu\nu}\, {\rm Re}\,D^{\mu\nu}(\omega,q)\,{\rm
Im}\,V^{\nu\mu}_{A,\parallel }(\omega,q). \label{ColK}
\ee
In the spinless model, the kernel ${\cal K}(\omega)$ is determined by processes
described by the Feynman diagrams in Fig.~\ref{Scattering}a at $\gamma\ll\omega
\ll \alpha T_1$ and Fig.~\ref{Scattering}b at $\alpha T_1 \ll \omega \ll T_1$. In
the spinful case, collisions between electrons of the same chirality but opposite
spin (Fig.~\ref{Scattering}c), give the main contribution to ${\cal K}(\omega)$
for all $\omega \ll T_1$. The frequency dependence and the asymptotes of ${\cal
K}(\omega)$ are shown in Fig.~\ref{Kw_plot}.

{\it Large energy transfer, $\omega \gg T_1$}.---In this case, we consider the
contributions to $I^{\mu\nu}= I_p^{\mu\nu}+I_s^{\mu\nu}+I_{eh}^{\mu\nu}$ from the
collective (plasmon/spinon) and e-h poles separately, see
Figs.~\ref{Scattering}d-\ref{Scattering}f. Because of the splitting of the e-h,
plasmon, and spinon poles, the collision integral would be nonlocal~\cite{catelani05} if expressed
solely in terms of the e-h distribution functions $N^{\mu\nu}_\omega$ [see
Eq.~(\ref{Neh})]. By introducing the plasmon/spinon distribution functions
$n^{\mu}_b (\omega)$ we can express the emission rate of the bosonic excitations
($b=p,s$) in the local form ($v_p \equiv u,\, v_s \equiv  v_F$):
\be
I_b^{\mu\nu}(\omega) = \sum_{\alpha} \omega {\cal L}^{\mu\nu}_{\alpha, b}(\omega)
\left[1 + n^\alpha_b(\omega)\right]~,
\ee
where the collision kernels ${\cal L}$ have the scaling form ${\cal
L}^{\mu\nu}_{\alpha, b} = (\gamma/2
\omega^2) (v_F^*/v_b)\,{\cal
A}^{\mu\nu}_{\alpha, c}$ and the $\cal A$-factors read
\bea
{\cal A}^{\mu\mu}_{\pm\mu,p} &=& \left[\eta (v_F^* - v_F) + v_F \pm u\right] / \eta v_F^* , \\
{\cal A}^{\mu,-\mu}_{\pm,p} &=& -(v_F^* - v_F)/v_F^*, \quad
{\cal A}^{\mu\nu}_{\alpha, s} = (\eta-1) \delta_{\mu\nu} \delta_{\mu\alpha}.
\nonumber
\eea
In turn,  $n^{\mu}_b(\omega)$ satisfies the kinetic equation
\be
(\partial_t + \mu v_b\, \partial_x ) n^{\mu}_{b}(\omega)  =
 -\gamma_b n^{\mu}_{b}(\omega)  + \frac{\eta\gamma}{2}\sum_{\alpha\beta}
{\cal A}^{\alpha\beta}_{\mu,b}\,\, N^{\alpha\beta}_\omega,
 \label{Bosons_KE}
\ee
where $\gamma_b = \gamma ( v_F/v_F^*)$ and we used the relation $D^{\mu\nu}_K \simeq
2 \,{\rm Re} D^{\mu\nu}_R (1 + 2 N_\omega^{\mu\nu})$. This kinetic equation
describes the decay/creation of plasmons and spinons in/from {\it e-h\,} pairs.
Finally, the e-h pole in the emission rate (\ref{KernK}) gives
$I_{eh}^{\mu\nu}(\omega) = \delta_{\mu\nu}(\gamma/\omega) N^{\mu\mu}_\omega$. Note
that the collision kernels ${\cal L}^{++}_{+,b}\simeq {\cal L}_{eh} =
\gamma/\omega^2$ do not contain $\alpha$ as a small parameter. 
These most efficient relaxation processes involve electrons 
and bosons within the same chiral branch: their rates are resonantly enhanced
since $v_b$ and $v_F^*$ are close to each other. As a result, $\alpha$
appears in the combination $\alpha^2/|v_F^*-v_b|^\eta\sim 1$. It can also be shown
that the total kernel, ${\cal L} = {\cal L}_{eh} + (1/2)\sum_{\mu\nu\alpha,b}{\cal
L}^{\mu\nu}_{\alpha,b}$, is equal to the asymptotic value of ${\cal K} (\omega)$
[Eq.~(\ref{ColK})] in the limit $\omega \gg T_1$.

\begin{figure}[t]
\includegraphics[width=2.5in]{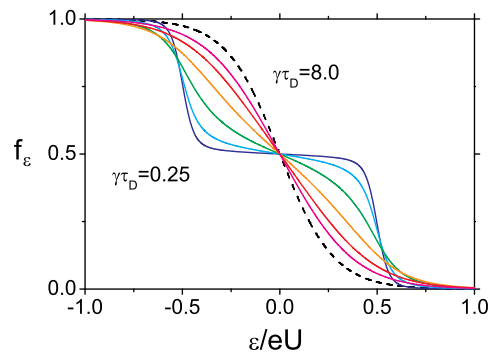}
\caption{
Electron distribution function
in the middle of a quantum wire of length $L$ biased by $eU=40 T$, for
$T=T_1/2$. The solid lines correspond
to $\gamma\tau_D$ = 0.25, 0.5, 1, 2, 4, and 8, where $\tau_D=L/v_F^*$ and
$\gamma$ is the elastic backscattering rate.
The dashed curve shows the limiting Fermi distribution.}
\vspace{-3mm}
\label{disfun}
\end{figure}

{\it Energy relaxation rate}.---We now turn to the
energy-relaxation rate $\gamma_E$ in the limit of weak nonequilibrium. Linearizing
the electron and boson kinetic equations, we estimate $\gamma_E$ from the total
\cite{Comment} collision kernel (Fig.~\ref{Kw_plot})
\be
{\gamma_E}\!\sim \!T^{-1}\int_0^T\!\!\!d\omega\,\omega^2{\cal K}(\omega)\sim
T^2{\cal K}(T). \label{Tau_E}
\ee
%where $T$ is understood as $\max\{T,eU\}$.
The characteristic $\omega$ for
$\gamma_E$ is of order $T$. Since ${\cal K}(\omega)\simeq 2\gamma/\omega^2$ for
$\omega\gg T_1$, $\gamma_E$ at $T\gg T_1$ does not depend on $\alpha$ and is given by the
backscattering rate:
\be
\gamma_E\sim\gamma~,\quad T\gg T_1 \label{main}
\ee
($\alpha$ enters only through the condition on $T$, so that the
result is also valid for the long-range Coulomb interaction).
At $T \alt T_1$, Eq.~(\ref{main}) is only valid for short wires of length $L \alt
v_F/\gamma$---otherwise the system is localized---and shows that a full
equilibration then has no time to develop. Note that $\gamma_E$ for spinful and
spinless electrons turn out to be parametrically the same, in contrast to the
weak-localization phase-relaxation rate $\gamma_\phi$ \cite{yashenkin08}. 

%Naively,
%one could expect that in the presence of spin the energy relaxation is
%enhanced---similarly to $\gamma_\phi$---due to strong scattering between electrons
%of the same chirality \cite{gornyi05,yashenkin08}.

In the limit of strong nonequilibrium, we solve Eqs.~(\ref{Fermions_KE}),
(\ref{Bosons_KE}) numerically to obtain the distribution function of electrons
$f_\epsilon$ in a wire biased by a voltage $U\gg T/e$, where $T$ is the
temperature in the leads, as a function of the distance to the contacts
($f_\epsilon$ shows up directly in tunneling spectroscopy \cite{chen09}; for
experiments on multi-channel wires, see, e.g., Ref.~\cite{pothier97}). The result,
shown in Fig.~\ref{disfun}, confirms the estimate (\ref{main}): the scale of
$\tau_D=L/v_F^*$ on which $f_\epsilon$ equilibrates is $\gamma^{-1}$, despite
$\alpha\ll 1$. For small $\alpha$, the inelastic processes involving opposite
chiralites ($\cal L^{++}_{-}$ and $\cal L^{+-}_{+}$), as well as the
backscattering of plasmons on the boundaries \cite{oreg93,fazio98,remark2,gutman09}, can be
neglected: the curves in Fig.~\ref{disfun} are thus $\alpha$-independent. At
$\tau_D \gg \gamma^{-1}$, $f_\epsilon$ approaches the Fermi distribution with the
temperature $T_e=\sqrt{3}eU/4\pi$.

{\it Discussion.}---As follows from  Eqs.~(\ref{main}) and (\ref{Bosons_KE}), the
thermalization of electrons occurs on the same time scale as the lifetime of {\it
bosons}. This elucidates a conceptually important point: boson decay
is a source of the inelastic relaxation of electrons.
Indeed, in the homogeneous case, the combination ${\rm Re} D\,{\rm
Im}V$ in Eq.~(\ref{ColK}) yields zero energy transfer in the e-e scattering
\cite{gornyi05}. However, any plasmon scattering broadens the peak in ${\rm
Im} V$, thus allowing for a finite transfer (even if $D$ remains free---no electron
backscattering) \cite{remark2}. This
is true, in particular, in an inhomogeneous LL without impurities but with a
nonuniform interaction $\alpha (x)$ \cite{oreg93,fazio98,remark2,gutman09}.

In a disordered LL, the spectral function of dRPA bosons 
is characterized by the rate $\gamma_b\simeq \gamma$ due to
the elastic scattering off impurities~\cite{gornyi05}.
One sees from Eq.~(\ref{Bosons_KE}) that out
of equilibrium [when $n^\mu(\omega)\neq N^{\mu\mu}_\omega$] the boson scattering
is represented entirely as the creation/annihilation of e-h pairs. As a result,
the inverse process---the inelastic electron scattering due to the emission/absorption
of bosonic excitations---is characterized by the same rate, $\gamma_E\sim\gamma$.
Impurities induce also the anharmonic decay of plasmons, as well as
their inelastic scattering on each other. These processes have been
neglected as higher-loop corrections to the dRPA: their rates are much smaller
than $\gamma_E$. It is the latter that gives the thermalization rate for the full
bosonic distribution function $n(\omega)$ in Eq.~(\ref{Bosons_KE}) due to the
coupling (through $N_\omega$) to the fermionic function $f_\epsilon$. 

{\it Conclusion.}---We have formulated the analytical framework for
disordered LLs out of equilibrium, based on the kinetic equations for the boson and
electron distribution functions. We have found the equilibration rate, which,
remarkably, coincides with the elastic scattering rate. The kinetic approach
developed here is particularly convenient for studying heat transport and current
noise in strongly-correlated disordered 1D systems \cite{remark_future}.

We thank A. Altland, N. Birge, D. Gutman, A. Kamenev, M. Kiselev, J. Meyer, A.
Mirlin, Y. Nazarov, H. Pothier, A. Yashenkin, and I. Yurkevich for interesting
discussions. The work was supported by EUROHORCS/ESF (Project ``Quantum
Transport in Nanostructures"), by GIF Grant No.\ 965, and by CFN/DFG.


\begin{thebibliography}{99}
\bibitem{remark_ghost} In contrast to Ref.~\cite{catelani05},
    and also Refs.~\cite{bagrets08,gutman08},
    we construct our theory without the so-called ``ghost modes".
\bibitem{Comment} A complementary approach has been formulated for spinless
    electrons in
    Ref.~\cite{bagrets08} in terms of ``undressed" fermions, which gives the
    same result for ${\cal K}(\omega)$.
\bibitem{remark2} In a finite-length LL with, e.g., $\alpha
    (x)\propto \theta(x)\theta (L-x)$, plasmons are backscattered off the noninteracting
    contacts. Inhomogeneities in the interaction strength induce the inelastic electron
    scattering \cite{oreg93} as well.
    This type of inelastic relaxation follows directly
    from the kinetic equations (\ref{Fermions_KE}),(\ref{Bosons_KE}) with the kernels
    (\ref{KernK}),(\ref{ColK}) taken for the space-dependent $\alpha (x)$.
    In a different framework---by directly calculating the Keldysh Green's function---the
    relaxation of the electron distribution in a finite LL has also been found in Ref.~\cite{gutman09}.
    This mechanism of evolution of $f_\epsilon$ combines with the impurity-induced equilibration.
\end{thebibliography}
\end{document}